\documentclass{PoS}

\usepackage{upgreek}
\usepackage{amsmath}
\usepackage{url}
\usepackage{xspace}
\usepackage{color}

\def\lsi{LS~I~+61~303\xspace}
\def\g{$\upgamma$\xspace}

\title{Review on the multiwavelength emission of the gamma-ray binary LS~I~+61~303}

\ShortTitle{Review on the multiwavelength emission of the gamma-ray binary LS~I~+61~303}

\author{\speaker{Benito Marcote}\\
        Joint Institute for VLBI ERIC (JIVE)\\
        E-mail: \email{marcote@jive.eu}}

\abstract{
    Gamma-ray binaries are systems composed of a massive star and a compact object that produce emission from radio to very high energy \g-rays. LS~I~+61~303 is one of the only six gamma-ray binaries discovered so far. It is thought that gamma-ray binaries contain a young highly rotating neutron star as compact object, and the emission is produced by the interaction between its relativistic pulsar wind and the stellar wind, However, in the case of \lsi a microquasar scenario is still considered and results pointing to oppose directions have been published during the last decades. Here we provide a review about the state of the art of \lsi, summarizing the observed emission from radio to very high energy \g-rays along all these years, and we discuss the proposed scenarios that can explain such emission.
}

\FullConference{XII Multifrequency Behaviour of High Energy Cosmic Sources Workshop\\
		12-17 June, 2017\\
		Palermo, Italy}

\begin{document}

\section{Introduction} \label{sec:intro}

Only a reduced number of binary systems has shown high-energy (HE; $0.1\text{--}100\ \mathrm{GeV}$) and/or very high-energy (VHE; $\gtrsim 0.1\ \mathrm{TeV}$) $\upgamma$-ray emission so far. This requires the presence of powerful mechanisms that accelerate particles up to relativistic energies, making these systems unique to study in detail particle acceleration given the short timescales involved and their relative proximity.

During the last decades different types of binaries have been discovered as non-pulsed HE emitters: three high-mass X-ray binaries (Cyg~X-1, Cyg~X-3, SS~433), several neutron-star/low-mass star binaries, one colliding wind binary ($\upeta$-Car), and six gamma-ray binaries. Only the latter group displays significant VHE emission (see \cite{dubus2013,dubus2015} and references therein).

All the known gamma-ray binaries are composed of a high-mass star (of either O or B spectral type) and a compact object which only in one case (PSR~B1259$-$63) has been confirmed to be a neutron star. In all the other systems the nature of this compact object remains unclear and could be either a black hole or a neutron star. In addition to the presence of VHE emission, gamma-ray binaries are characterized by displaying a Spectral Energy Distribution (SED) dominated by the $\upgamma$-ray photons.
This behavior makes gamma-ray binaries unique among all binaries displaying $\upgamma$-ray emission, as for all the other kinds of binaries the SED is clearly dominated by the X-ray photons. A different origin for the emission is thought to be underlying this different behavior.

In this work we present a review on the gamma-ray binary LS~I~+61~303. We present the source in Sect.~\ref{sec:intro-lsi}. We describe the multiwavelength emission in Sect.~\ref{sec:emission} and the proposed scenarios to explain the collected data in Sect.~\ref{sec:models}. Finally, we present the conclusions in Sect.~\ref{sec:conclusions}.

\section{The gamma-ray binary system LS~I~+61~303} \label{sec:intro-lsi}

The gamma-ray binary LS~I~+61~303 is composed of a young, rapidly rotating, $10$--$15~\mathrm{M_{\odot}}$ B0~Ve star \cite{hutchings1981} and a compact object orbiting it every $P_\text{orb} = 26.496 \pm 0.003\ \text{d}$ \cite{gregory2002} in an eccentric orbit ($e = 0.72 \pm 0.15$). For reference, the periastron passage occurs at an orbital phase of $\phi_{\rm orb} = 0.23 \pm 0.03$ \cite{casares2005lsi61303}. The whole system is located at a distante of $2.0 \pm 0.2\ \text{kpc}$ according to {\sc HI} measurements \cite{frail1991}.
The nature of the compact object remains unclear, and can be either a neutron star (NS) or a stellar-mass black hole (BH). The former one is favored for high inclinations of the orbit ($\gtrsim 25^{\circ}$), whereas the latter one is favored otherwise. We note that inclinations above $60^{\circ}$ are already discarded according to the orbital parameters \cite{casares2005lsi61303,aragona2009}.
Whereas pulse searches have been conducted, no pulsed emission have been detected so far \cite{mcswain2011,canellas2012}. This however does not reject the presence of a pulsar as we would expect the pulses to be absorbed due to the compactness of the system, as it occurs in the much wider orbit of PSR~B1259$-$63 during periastron.

On the one hand, the mass of the compact object has been estimated to be $1.3\ \mathrm{M_\odot} < M_\text{c} < 2.0\ \mathrm{M_\odot}$ (implying the existence of a a neutron star) assuming that the orbit and the Be decretion disk are in the same plane, and the inclination of the orbit with respect to the observer is $\sim 65^\circ < i < 75^\circ$ \cite{zamanov2017}. Given the uncertainties in the measurement of the inclination, larger values of the mass are actually possible, allowing the existence of a black hole.
On the other hand, the existence of a black hole has been suggested by indirect studies on the X-ray luminosity and photon index \cite{massi2017}.

The companion star has a circumstellar disk that extends over periastron, causing a direct interaction with the compact object during periastron, where the disk could be disrupted.
Although the shape of the disk remains unclear, an elliptical shape \cite{xing2017} or one-armed spiral density wave \cite{negueruela1998,paredesfortuny2015} have been suggested. In any case, it is clear that the size of the disk changes along the so-called superorbital modulation (see next section) \cite{zamanov2000,zamanov2013}.

\section{Multiwavelength emission} \label{sec:emission}

LS~I~+61~303 exhibits persistent and orbitally modulated emission from radio to VHE \g-rays. Whereas the first HE counterpart was already detected by {\em COS~B} \cite{gregory1978}, its association could not be confirmed until the discovery of a variable VHE source, almost 30~yr later \cite{albert2006}.


A long-term modulation have also been found at all wavelengths. The so-called superorbital modulation exhibits a period of $P_\text{so} = 1\,667 \pm 8\ \text{d}$, or $\sim 4.6\ \text{yr}$ \cite{gregory1989,paredes1990,gregory2002} and affects the amplitudes of the 26.5-d bursts but also the phases at which these outbursts take place (e.g. the peak of the radio bursts takes place at orbital phases shifting from 0.45 to 0.95 along the superorbital period).

In the following, we summarize the observed emission at different wavelengths, from VHE \g-rays to radio frequencies.

\subsection{VHE \g-ray emission}

LS~I~+61~303 has been detected at VHE (TeV energies) with the MAGIC Cherenkov Telescopes \cite{albert2006} and with VERITAS \cite{acciari2008}. This emission show a modulation coincident with the orbital period \cite{albert2009} with outbursts peaking at orbital phases $\phi_{\rm orb} \sim 0.6$--$0.8$.
The emission outside the outbursts (e.g. at $\phi_{\rm orb} \sim 0.0$) is faint and thus difficult to detect, though it has been reported with more sensitive observations \cite{acciari2011}. Therefore it is thought that there is persistent VHE emission at all orbital phases.
This emission is explained by a power-law with a photon index of $\Gamma \approx 2.4$--$2.7$ and a flux at 1~TeV of $2.4$--$2.9 \times 10^{-12}\ \mathrm{TeV^{-1}\ cm^{-2}\ s^{-1}}$ \cite{aleksic2012lsi61303}.

A long-term modulation has recently been reported at VHE \cite{ahnen2016} with a period of $1\,610 \pm 58\ \text{d}$, thus consistent with the superorbital modulation observed at other wavelengths. However, still more data are required to study in detail the behavior of LS~I~+61~303 at these energies.

\subsection{HE \g-ray emission}

LS~I~+61~303 is the 12$^\text{th}$ brightest source in the HE sky, and was detected as a variable source in the {\em Fermi}/LAT data \cite{abdo2009} with a modulation consistent with the orbital period \cite{hadasch2012}.
The HE light-curve is roughly anti-correlated with respect to the VHE, X-ray, and radio ones, with the maxima taking place right after periastron.
The HE spectrum is described by a power-law with a photon index of $\Gamma \approx 2.1$ and a flux at 0.1~GeV of $9.5 \times 10^{-7}\ \mathrm{cm^{-2}\ s^{-1}}$ \cite{hadasch2012}. An exponential cutoff is observed at 3.9~GeV, though at higher energies (above 30~GeV) a new component emerges.

The HE emission also exhibits superorbital modulation \cite{ackermann2013,xing2017}, and a significant dip around the periastron has been reported at particular superorbital phases \cite{xing2017}.

\subsection{X-ray emission}

\lsi exhibits a non-thermal X-ray emission, which is also orbitally modulated and exhibits outbursts between orbital phases 0.4 and 0.8 \cite{paredes1997,harrison2000,chernyakova2012}. A correlation between the X-ray light-curve and the TeV one has been suggested but not confirmed \cite{anderhub2009,acciari2009}.
Rapid activity, with flares lasting for tens to hundreds of seconds, has been reported \cite{smith2009,li2011}, although quasi-periodic oscillations have been discarded. The superorbital modulation is also clearly observed at X-rays with a sinusoidal evolution \cite{li2012,chernyakova2012,chernyakova2017}, but a shift of $280 \pm 44$~d with respect to the radio modulation is reported (we note the period of 1\,667~d).

The X-ray emission is well-described by an absorbed power-law with an index of $\Gamma = 1.5$--$1.9$ and fluxes of $0.5$--$3 \times 10^{-11}\ \mathrm{erg\ s^{-1}\ cm^{-2}}$ \cite{paredes2007}. No significant intrinsic absorption is however reported, and the spectrum does not reveal signatures of accretion, absorption and/or emission lines \cite{sidoli2006}.
It has been reported the existence of a correlation between the flux and photon index, with a harder spectrum at higher fluxes \cite{smith2009,li2011}.

Finally, hints of X-ray extended emission at scales of $\sim 1$~arcsec have been claimed in the past \cite{paredes2007}, similar to the ones recently reported in PSR~B1259$-$63 that are ejected after periastron passage and move away from the system at $\sim 0.1c$ \cite{pavlov2015}.

\subsection{Optical emission}

The optical light-curve of \lsi is also orbitally \cite{hutchings1981,mendelson1989} and superorbitally \cite{paredesfortuny2015} modulated.
Measurements of the equivalent width of the {\sc H$\upalpha$} ($EW_{\sc H\upalpha}$) emission line, which is a good tracer of the conditions in the outer circumstellar disk, also show the orbital and superorbital variability \cite{zamanov1999,mcswain2010,zamanov2013,paredesfortuny2015}.
The photometric and $EW_{\sc H\upalpha}$ light-curves show the same behavior but with a shift with respect to the radio and X-ray light-curves, suggesting a coupling between these different wavelengths, and thus between the thermal and non-thermal emission.

\subsection{Radio emission}

The spectrum of LS~I~+61~303 has been widely studied at gigahertz frequencies along the orbit \cite{paredes1990,ray1997,strickman1998,zimmermann2015} and along the superorbital modulation \cite{gregory2002}. Whereas an outburst is observed every orbital cycle, their shapes and peak times slightly change from cycle to cycle \cite{ray1997}. The bursts are known to peak roughly at the same times ($\phi_{\rm orb} \approx 0.4$--$0.9$) for frequencies 1--25~GHz, exhibiting an average spectral index of $\alpha \approx -0.5$ that can smoothly be increased up to $\alpha \approx 0.0$ along the outbursts \cite{strickman1998,zimmermann2015}. The radio outbursts are delayed with respect to the X-ray outbursts with $\Delta\phi_{\rm orb} \approx 0.2$ \cite{chernyakova2012}.

Radio observations at low frequencies ($\sim 100\text{\,--\,}700~\mathrm{MHz}$) have also been conducted \cite{pandey2007,marcote2016}. A significant delay in the peak of the outbursts and their shapes has been reported \cite{marcote2016}, probably due to absorption. Additionally, these observations show evidences of a low-frequency turnover, to take place between 0.6 and 2~GHz.

Very long baseline interferometric (VLBI) radio observations show that the radio emission is resolved on milliarcsecond scales (5--10~AU) \cite{massi1993}. The structure shows periodic morphological changes along the orbit that have been interpreted as a signature of a cometary tail produced as a result of colliding winds \cite{dhawan2006,albert2008,moldon2012lsi61303,moldon2012thesis}. Although other interpretations of the data suggest the presence of a precessing jet as in a microblazar \cite{massi2012}.

\section{Microquasar or young non-accreting pulsar wind scenario?} \label{sec:models}


Different scenarios have been proposed to explain the aforementioned emission of \lsi and we still do not have a clear, full, picture. While a young highly rotating pulsar producing a shock between its relativistic wind and the stellar wind is thought to be present in all gamma-ray binaries (which would explain the significant differences observed with respect to the other \g-ray emitting X-ray binaries), in the case of \lsi a microquasar scenario is still considered.

In the following we summarize the proposed scenarios that has been proposed in the literature to explain the available data.

\subsection{Microquasar scenario}

In the microquasar scenario \cite{taylor1984} an accretion onto the compact object is required. This accretion would lead the formation of jets where the particles are accelerated and would emit up to VHE \g-rays due to upscattering of stellar photons \cite{boschramon2004}. This is the scenario observed in X-ray binaries, either including black holes or accreting powered neutron stars.

The reported anti-correlation between the X-ray flux and the photon index is similar to the one observed in other black-hole X-ray binaries, but not in PSR~B1259$-$63, which has leaded to the suggestion of the existence of a black hole in \lsi \cite{massi2017}. Furthermore, through frequency radio studies it has been suggested that there are actually two close periods of $\approx 26.5$ and $\approx 26.9$~d modulating the emission of \lsi, and the superorbital modulation is just the beating period between them \cite{massi2014,massi2015,massi2016}. In this case, we would assume the presence of a precessing jet with a period close to the orbital period of the system, naturally explaining the regularity of the long-term emission \cite{jaron2016,saha2016}.

\subsection{Young non-accreting pulsar wind scenario}

To explain the vast differences observed in the SED between the gamma-ray binaries and the \g-ray emitting X-ray binaries a different underlying scenario was proposed, which has been confirmed at least for the only gamma-ray binary with a confirmed pulsar, PSR~B1259$-$63.
This scenario requires the presence of a neutron star, which must be young and must be highly rotating to guarantee to be rotation powered, and not accretion powered. The neutron star thus produces relativistic winds that collide with the stellar wind originating a strong shock due to the proximity of both components \cite{maraschi1981}. In this case particles are accelerated at the interaction region between both winds but also in the Coriolis turnover behind the compact object \cite{dubus2006}.

\lsi hosts a B spectral-type star with a circumstellar disk which is perturbed by the compact object during periastron. In this sense we would expect a system qualitatively similar to PSR~B1259$-$63, but in a much closer orbit.
Indeed, we observe an outburst per orbital cycle that is produced after the compact object crosses the circumstellar disk. Given the lack of evidences of accretion in the system, a wind interaction is favored in \lsi.

It has been suggested that the superorbital variability is related to periodic changes in the mass-loss rate of the Be star and/or variations in the circumstellar disk \cite{mcswain2010,paredesfortuny2015}. Therefore this variability could not be linked with only the compact object (e.g. a precessing jet).
This connection is supported by the evidences of periodic changes in the circumstellar disk, with the presence of a spiral density wave \cite{mcswain2010} which has been interpreted as an non-uniform, elongated, disk \cite{xing2017} or with one-armed spiral shape \cite{paredesfortuny2015}. In any of these scenarios, the rotation of the disk with a long period would naturally produce the observed superorbital modulation. When the putative neutron star approaches periastron it would face a region of the disk with different density, producing a change in the strength of the interaction and emission each time.

Independently to this, we note the X-ray extended emission that has been reported in LS~I +61~303 \cite{paredes2007}. This phenomena could be related to the similar emission observed in PSR~B1259$-$63 \cite{pavlov2015}. This emission has been interpreted as material expelled from the system after the periastron passage, and it would be expected in systems with eccentric orbits with a neutron star disrupting the circumstellar disk \cite{barkov2016}. The orbit of LS~I~+61~303 achieves these conditions, so new detailed X-ray observations to confirm and study this extended X-ray emission would provide a strong argument towards the existence of a neutron star.

\subsection{Flip-flop model}

A third scenario, the so-called flip-flop model, has been proposed to explain the emission observed in \lsi \cite{papitto2012,torres2012}. This scenario also considers the presence of a neutron star as compact object. However, in this case the neutron star would change state from a propeller regime (during periastron) to an ejector region (at apastron). These changes of state would be driven by the interaction with the circumstellar disk and the stellar wind. At periastron the neutron star magnetosphere would be compressed and disrupted due to the much denser surrounding of the disk, suppressing the synchrotron emission and favoring the propeller regime.
At apastron, far from the circumstellar disk, the neutron star would change again to the ejector regime favoring the strong VHE emission that is observed.

\section{Conclusions} \label{sec:conclusions}

LS~I~+61~303 is one of the six known gamma-ray binaries. Whereas it is thought that all gamma-ray binaries share a common scenario, different from the microquasar scenario observed in X-ray binaries, in the case of \lsi this scenario is still considered.

\lsi exhibits a SED dominated by the HE photons, as all the other gamma-ray binaries, and in contrast to the ones observed in \g-ray emitting X-ray binaries, where the X-ray photons dominate due to the presence of accretion and a strong cutoff is observed above those energies.
The stablished connection between changes in the thermal emission (arising from the circumstellar disk) and the non-thermal emission (from the interaction of the compact object and the disk/star), among other phenomena, points to an emission driven by wind collisions instead of jet-related.

The nature of the compact object would allow us to unveil easily the kind of interaction that takes place in \lsi. However, we note the existence of indirect evidences pointing to opposed directions. Therefore only a direct measurement of the mass of the compact object would firmly unveil its nature.

\section*{Acknowledgements}

The author acknowledges support from the Spanish Ministerio de Econom\'ia y Competitividad (MINECO) under grants AYA2016-76012-C3-1-P and MDM-2014-0369 of ICCUB (Unidad de Excelencia ``Mar\'ia de Maeztu'').

\bibliographystyle{JHEP}
\bibliography{/home/marcote/Documents/Reference/bibliography.bib}

\bigskip
\bigskip
\noindent {\bf DISCUSSION}

\bigskip                     
\noindent {\bf AGNIESZKA JANIUK:} Among the six $\upgamma$-ray binaries, what source would be the best candidate to contain a black hole instead of a neutron star?
What would be the scenario for $\upgamma$-ray production in case of a such BH/HMXB system?

\bigskip                     
\noindent {\bf BENITO MARCOTE:}  We only know the presence of a neutron star in one system, PSR~B1259$-$63, and there are strong evidences in another one, 1FGL~J1018.6$-$5856. For the other three well-studied gamma-ray binaries (the extragalactic LMC~P3 remains widely unexplored), it is thought that the compact object is a neutron star too, although as we have mentioned, \lsi has been claimed to contain a black hole in some interpretations. In this case the $\upgamma$-ray production should arise from upscattering of stellar photons from the relativistic particles accelerated in the jets.

\end{document}